\pdfoutput=1
\pdfoutput=1
\pdfoutput=1
\pdfoutput=1
\pdfoutput=1
\documentclass[journal=jacsat,manuscript=article]{achemso}
\usepackage{graphicx}
\usepackage{amsmath}
\usepackage{subfig} 
\usepackage{booktabs}
\usepackage{expl3}
\usepackage{amsmath}
\usepackage{subfig} 
\usepackage[version=3]{mhchem} 

\author{Fardin Ghorbani}
\email{fardin.ghorbania@gmail.com}
\author{Sina Beyraghi}
\author{Javad Shabanpour}
\email{m.javadshabanpour1372@gmail.com}
\author{Homayoon Oraizi}
\author{Hossein Soleimani}
\author{Mohammad Soleimani}

\affiliation[Iran University of Science and Technology]
{School of Electrical Engineering, Iran University of Science and Technology, Narmak, Tehran 16486-13114, Iran}

\title[An \textsf{achemso} demo]
{Deep neural network-based automatic metasurface design with a wide frequency range}

\abbreviations{IR,NMR,UV}

\begin{document}
	\begin{abstract}
Beyond the scope of conventional metasurface which necessitates plenty of computational resources and time, an inverse design approach using machine learning algorithms promises an effective way for metasurfaces design. In this paper, benefiting from Deep Neural Network (DNN), an inverse design procedure of a metasurface in an ultra-wide working frequency band is presented where the output unit cell structure can be directly computed by a specified design target. To reach the highest working frequency, for training the DNN, we consider 8 ring-shaped patterns to generate resonant notches at a wide range of working frequencies from 4 to 45 GHz. We propose two network architectures. In one architecture,  we restricted the output of the DNN, so the network can only generate the metasurface structure from the input of 8 ring-shaped patterns. This approach drastically reduces the computational time, while keeping the network's accuracy above 91\%. We show that our model based on DNN can satisfactorily generate the output metasurface structure with an average accuracy of over 90\% in both network architectures. Determination of  the metasurface structure directly without time-consuming optimization procedures, having ultra-wide working frequency, and high average accuracy equip an inspiring platform for engineering projects without the need for complex electromagnetic theory.  
	\end{abstract}
	
	\section{1. Introduction}
Metamaterials,as artificial media composed of engineered subwavelength periodic or nonperiodic geometric arrays, have witnessed enormous attentions due to their exotic properties to modify the permittivity and permeability of materials\cite{pendry2000,rajabalipanah2019asymmetric,shabanpour2020full}. Today, just two decades after the first implementation of metamaterials by David Smith and colleagues\cite{smith2000composite} who unearthed Veselago’s original paper\cite{veselago1967electrodynamics}, metamaterials and their 2D counterpart, metasurfaces, have been widely used in practical applications such as, but not limited to, polarization conversion\cite{grady2013terahertz,shabanpour2101implementation}, reconfigurable wave manipulation\cite{shabanpour2020programmable,hashemi2016electronically}, vortex generation\cite{yang2014dielectric,shabanpour2020ultrafast}, and perfect absorption\cite{landy2008perfect,shabanpour2020reconfigurable}.  

However, all of the above-mentioned works are based on a traditional design approaches, consisting of model designs, trial-and-error method, parameter sweep, and optimization algorithms. Conducting numerical full-wave numerical simulations assisted by optimization algorithm is a time-consuming process which consumes plenty of computing resources. Besides, if the design requirements change, simulations must be repeated afresh which impedes users from paying attention to their actual needs. Therefore, to fill the existing gaps to find a fast, efficient, and automated design approach, machine learning has been into our consideration.

Machine learning and its specific branch, deep learning, are an approaches to automatically learn the connection between input data and target data from the examples of past experiences. Machine learning is an effort to employ algorithms to devise a machine to learn and operate without explicitly planning and dictating individual actions. To be more specific, machine learning equips an inspiring platform to deduce the fundamental principles based on previously given data thus, for another given input, machines can make logical decisions automatically. With the ever-increasing evolution of machine learning and its potential capacity to handle crucial challenges, such as signal processing \cite{ghorbani2020eegsig}, and physical science \cite{carleo2019machine}, we are witnessing their applications to electromagnetic problems. Due to its remarkable potentials such as providing less computational resources, more accuracy, less design time, and more flexibility, machine learning has been entered in various wave-interaction phenomena, such as Electromagnetic Compatibility (EMC)\cite{medico2018,shi2020}, Antenna Optimization and Design \cite{cui2020,sharma2020}, All-Dielectric Metasurface\cite{an2019deep}, Optical and photonic structures\cite{so2019designing}, and Plasmonic nanostructure\cite{malkiel2018plasmonic}.

Recently, T.Cui et al. have proposed a deep learning-based metasurface design method named REACTIVE, which is capable of detecting the inner rules between a unit-cell building and its EM properties with an average accuracy of 76.5\% \cite{qiu2019deep}. A machine-learning method to realize anisotropic digital coding metasurfaces has been investigated, whereby 70000 training coding patterns have been applied to train the network\cite{zhang20199}. In Ref\cite{shan2020coding} a deep convolutional neural network has been studied to encode the programmable metasurface for steered multiple beam generation with an average accuracy of more than 94 percent. A metasurface inverse design method using a machine learning approach has been introduced in\cite{shi2020metasurface} to design an output unit cell for specified electromagnetic properties with 81\% accuracy in a low-frequency bandwidth of 16-20 GHz. Recently, a double deep Q-learning network (DDQN) to identify the right material type and optimize the design of metasurface holograms has been developed\cite{sajedian2019double}.

In this paper, benefiting from Deep Neural Network (DNN), an inverse design procedure of a metasurface with an average accuracy of up to 92 percent has been presented. Unlike the previous works, to reach the highest working frequency, we consider 8 ring-shaped digital distributions (See top left pf the \textbf{Fig. 1}) to generate resonant notches in a wide range of working frequencies from 4 to 45 GHz. Therefore, after training the deep learning model by a set of samples, our proposed model can automatically generate the desired metasurface pattern, with four predetermined reflection information (as number of resonances, resonance frequencies, resonance depth, and resonance bandwidths) for ultra-wide working frequency bands. Comparison of  the output of numerical simulations with the design target illustrates that our proposed approach is successful in generating corresponding metasurface structures with any desired S-parameter configurations. Determination of the metasurface structures directly without ill-posed optimization procedures, consuming less computational resources, ultra-wide working frequency bands, and high average accuracy paves the way for our approach to become beneficial for those engineers who are not specialists in the field of electromagnetic, thus, they can focus on their practical necessitates which significantly boost the speed of the engineering projects.   
 \section{2. METHODOLOGIES}
\subsection{2.1. Metasurface Design}

\textbf{Fig. 1} shows the schematic representation of the proposed metasurface structure consisting of three layers, from top to bottom, as a copper ring-shaped pattern layer, a dielectric layer, and a ground layer to impede the backward transmission of EM energy. FR4 is chosen as the substrate with permittivity of 4.2+0.025i, and thickness of h=1.5mm.  The top metallic layer comprises 8 ring-shaped patterns distributed side by side, each of which can be divided into 8 × 8 lattices labeled as “0” and “1” which denote
the area without and with the copper. Each metasurface composed of an infinite array of unit-cells. Each unit-cell consists of 4 × 4 randomly distributed of 8 × 8 ring-shaped patterns. Therefore, each unit cell comprises 32 × 32 lattices. The length of the lattices, periodicity of unit cells, and thickness of the copper metallic patterns are l = 0.2 mm, p = 6.4 mm, and t =0.018 mm respectively. Unlike the previous works\cite{qiu2019deep,shi2020metasurface}, defining 8 ring-shaped patterns to train the DNN is the novelty employed here to generate the desired resonance notches in a wide frequency band. Each ring-shaped pattern is designed in such a way to generate resonant notches at different frequencies from 4 to 45 GHz, thus, we can import the data set of S-parameters to train the network for our specified targets. It is almost impossible to obtain the relationship between the metasurface matrices and S-parameters. Due to the close connection between the metasurface pattern matrix and its corresponding reflection characteristics, the deep learning algorithm is used to  reduce the computational burden for obtaining the optimal solution.

	\begin{figure*}[t!]
	\centering
	\includegraphics[height=8cm]{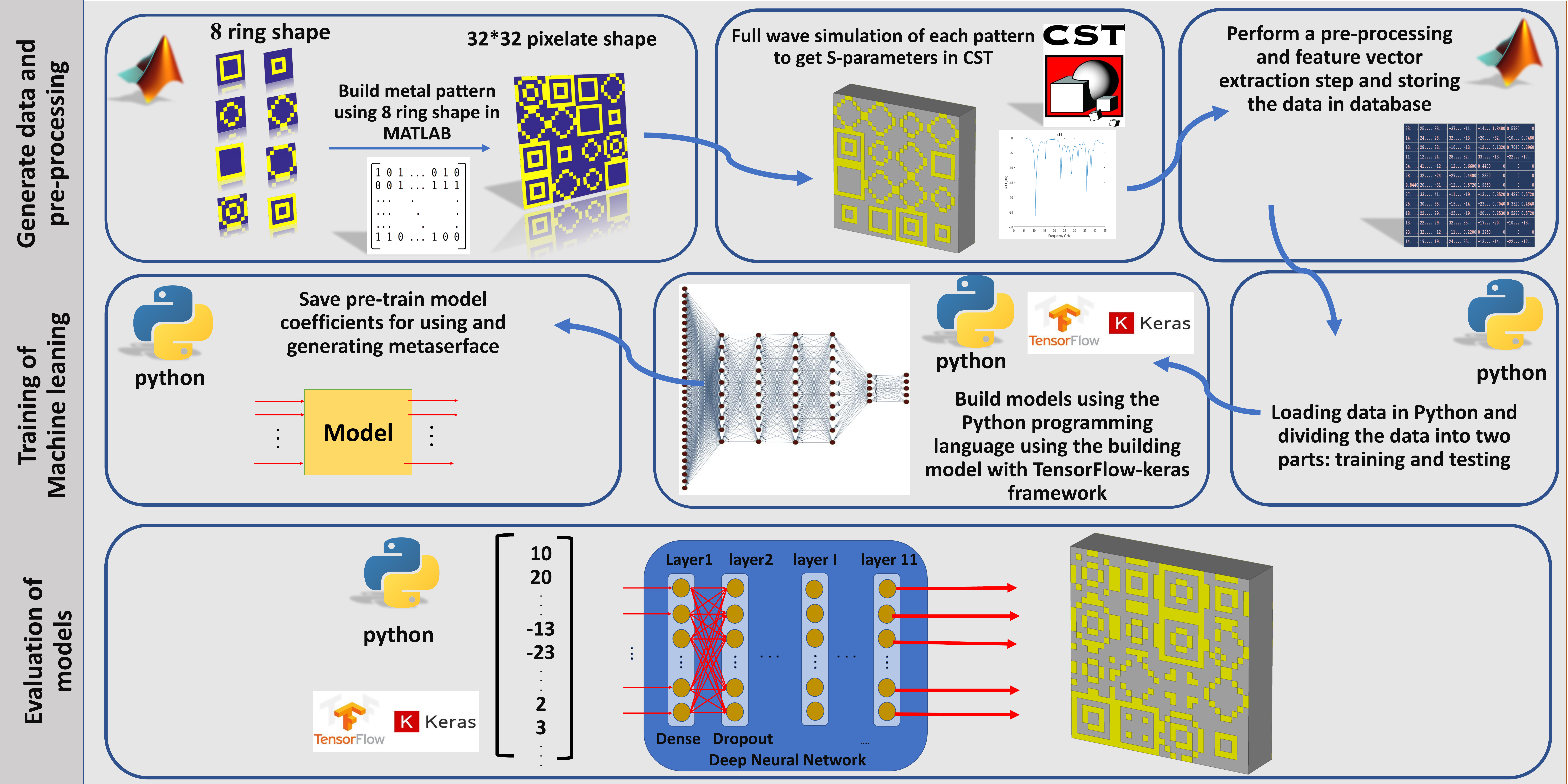}
	\caption{Sketch representation of the design process of DNN-based approach for metasurface inverse design. The process consists of three steps of generating data and pre-processing, Training of machine learning, and evaluations of a model.}
	\label{fgr:example2col}
\end{figure*}

\subsection{2.2. Deep Learning}

Artificial neural networks have emerged in the last two decades with many applications, especially in "optimization" and "artificial intelligence". \textbf{Fig. 2} shows an overview of an artificial neuron, with $X_{1}$, $X_{2}$, ... as its inputs (input neurons) . In neural networks, each $X$ has a weight, denoted by $W$. Observe that each input is connected to a weight; thus, each input must be multiplied by its weight. Then in the neural network, the sum function (sigma) adds the products of $X_{i}$'s by $W_{i}$'s. Finally, an activation function determines the  output of these operations. Then the output of neurons by the activation function $\phi(u)$, with b as a bias value is:
\begin{equation}
Y = \phi(\sum\limits_{i=1}^n W_{i}X_{i}+b_{i})
\end{equation}

\begin{figure*}[h]
	\centering
	\includegraphics[scale=0.1]{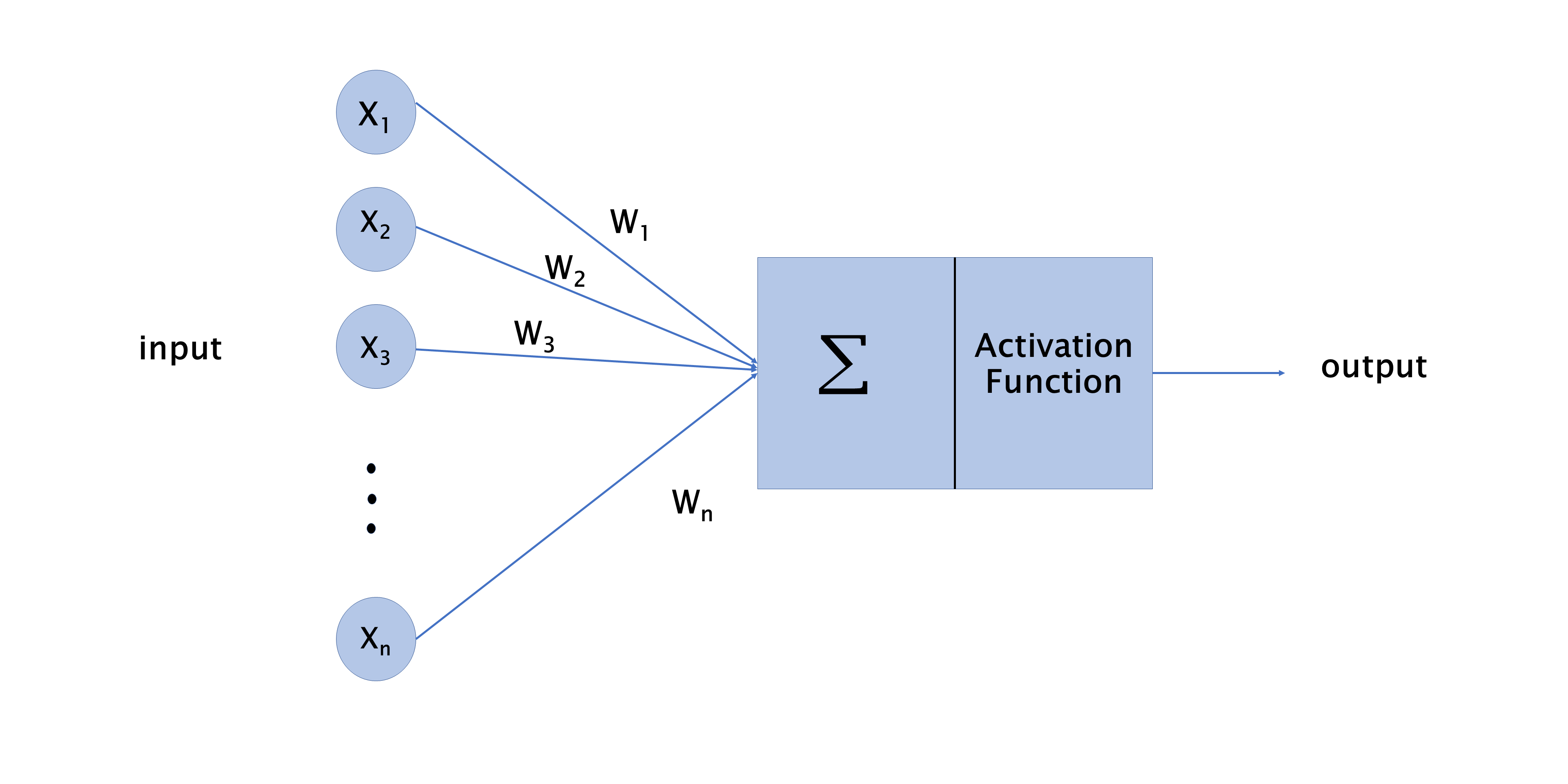}
	\caption{An overview of an artificial neuron} 
\end{figure*}
The neural network is made up of neurons in different layers. In general, a neural network consists of three layers: input, hidden, and output. The greater the number of layers and neurons in each hidden layer is, the more complex the model becomes. When the numbers of hidden layers and the number of neurons increases, our neural network becomes a deep neural network. In this work, we use a DNN to design the desired metasurface.  

\subsubsection{\textit{A. Non-restricted output}}

The inverse design of the metasurface is anticipated to determine the intrinsic relationships between the final metasurface structure and its geometrical dimensions by DNN. We have generated 2000 sets of random matrices that represent the metasurface structures using the “RAND” function
in MATLAB software. In the next step, we have linked the MATLAB with CST MWS to calculate the S-parameters of the metasurface. To calculate the reflection characteristics of the infinite arrays of the unit cells, we have conducted the simulations when the unit-cell boundary conditions are employed in x and y directions and open boundary conditions in the z-direction. 
Finally, when it comes to the design procedure, we only need to enter the predetermined EM reflection properties, and our model can generate the output metasurface based on the learned data during the training step. The dataset is established to generate  16 random numbers between 1 and 8 to form 4×4 matrices where each number represents one of the 8 ring-shaped patterns. To form our datasets, we have generated two thousand pairs of S-parameter and metasurface pattern matrices ( 70\% as a training set and 30\% as a testing set), and the output of the training model is a matrix of 32×32. Each unit-cell can generate 8 notches in the frequency band of 4 to 45 GHz. By defining three features for each resonance ( namely, notch frequency, notch depth, and notch bandwidth), the input of our proposed DNN  is a vector with dimension 24, and the output is a vector of dimension 1024, which represents a unit cell of 32 × 32 pixels. The details of the designed network are summarized in \textbf{Table 1}.

\begin{table}[h]
	\centering
	\caption{Detailed information of the non-restricted output network architecture.} 
	\scalebox{0.88}{
		\begin{tabular}{c|c|c|c|c}
			\hline
			Layer number & Layer                & output shape & number of parameter & activation function \\ \hline
			1            & dense\_1 (Dense)     & (None, 24)   & 600                 & relu                \\ \hline
			2            & dropout\_1 (Dropout) & (None, 24)   & 0                   & -                   \\ \hline
			3            & dense\_2 (Dense)     & (None, 300)  & 90300               & relu                \\ \hline
			4            & dropout\_2 (Dropout) & (None, 300)  & 0                   & -                   \\ \hline
			5            & dense\_3 (Dense)     & (None, 300)  & 90300               & relu                \\ \hline
			6            & dropout\_3 (Dropout) & (None, 300)  & 0                   & -                   \\ \hline
			7            & dense\_4 (Dense)     & (None, 300)  & 90300               & relu                \\ \hline
			8            & dropout\_4 (Dropout) & (None, 300)  & 0                   & -                   \\ \hline
			9            & dense\_5 (Dense)     & (None, 300)  & 90300               & relu                \\ \hline
			10           & dropout\_5 (Dropout) & (None, 300)  & 0                   & -                   \\ \hline
			11           & dense\_6 (Dense)     & (None, 1024) & 308224              & sigmoid             \\ \hline
	\end{tabular}}
	
\end{table}

In the proposed model, dense and dropout layers are used one after the other. In the fully connected (dense) layer, each neuron in the input layer is connected to all the neurons in the previous layers. In the dropout layer,  some neurons are accidentally ignored in the training process in order to avoid the misleading of the learning process as well as increasing the learning speed and reducing the risk of over-fitting. By selecting relevant features from the input data, the performance of the machine learning algorithms is efficiently enhanced.
In the proposed model, the values of batch size and learning rate are set to 30 and 0.001, respectively. Besides, the Adam optimization algorithm is used for tuning the weighting values ($W_{i}$). During the training process, the difference between original and generated data is calculated repeatedly by tuning and optimizing the weight values for each layer. When the difference reaches the satisfying predetermined criterion which is defined as loss function, then the training process stops. The Mean Square Error (MSE) is used as a loss function defined as:
\begin{equation}
{\rm{MSE}} = \frac{1}{N}\sum\limits_{i = 1}^N {{{({f_i} - {y_i})}^2}} 
\end{equation}
where $f_{i}$ and $y_{i}$ denote the anticipated value and the actual value,
respectively. For selecting an appropriate activation function, Since our desired output in the neural network is 0 or 1, we used the sigmoid function in the last layer, while using other activation functions reduce the accuracy. Formulation of the activation of relu and sigmoid functions are given  in equations 3 and 4, respectively:
\begin{equation}
\phi(x) = \begin{cases}
0 & x \leq 0 \\
x & x > 0
\end{cases} 
\end{equation}

\begin{equation}
\phi(x) = \dfrac{1}{1+e^{-x}}
\end{equation}

for validation, several design goals of S-parameters are suggested in anticipation that our proposed DNN is capable of producing equivalent unit-cell structures. The DNN algorithm is realized by the python version 3.8, and the Tensorflow and Keras framework\cite{chollet2015keras} are used to establish the model. As an example, a metasurface structure is designed with three notches using the DNN method. The specified reflection informations are [number of resonances; resonances frequencies; resonance depth; and the bandwidth of each resonance] = [ 3; 17.5, 23.5, 25.3 GHz; -30, -20, -20 dB; 0.5, 0.5, 0.4 GHz].Observe in \textbf{Fig. 3a}, that the output full-wave results achieve the design goals. 

For the next example, , a uni-cell is designed with one resonance frequency (-15 dB) at 15 GHz. The simulation result shows good conformity with our design target (See \textbf{Fig. 3b}). Furthermore, the curves of the mean square error and the accuracy of the presented non-restricted output DNN method are proposed in \textbf{Fig. 4} showing the accuracy rate higher than 92\%. 

	\begin{figure*}[h]
	\centering
	\includegraphics[height=4.4cm]{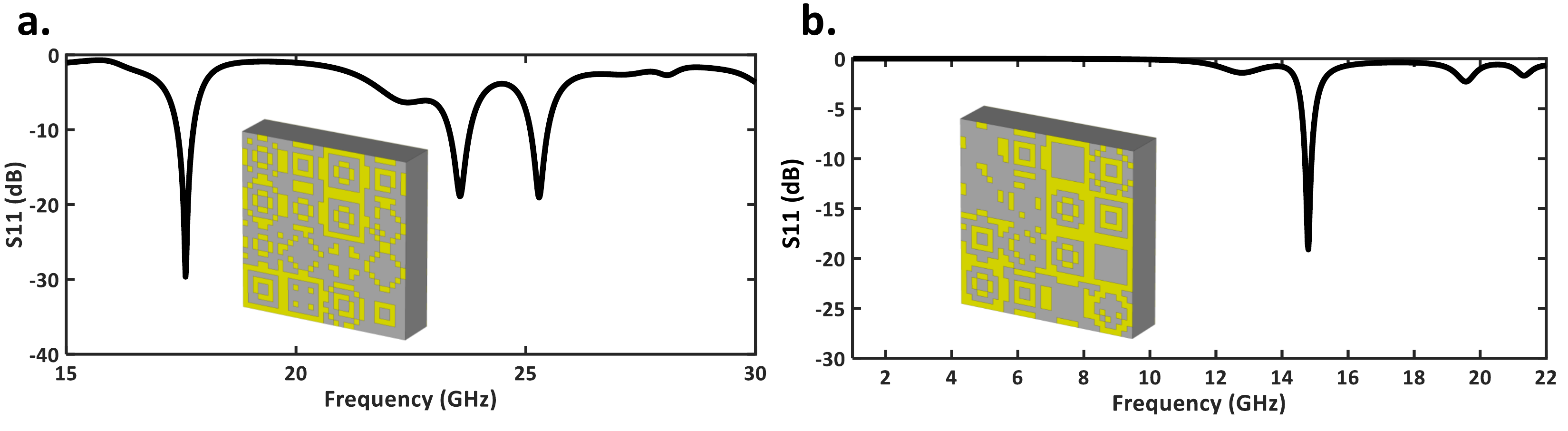}
	\caption{The simulated reflection coefficient of non-restricted output network architecture a) metasurface with three notches under -10 dB. b) metasurface with a single notch under -10 dB.}
	\label{fgr:example2col}
\end{figure*}
\begin{figure*}[h]
	\centering
	\subfloat[Accuracy]{\includegraphics[width=7.5cm]{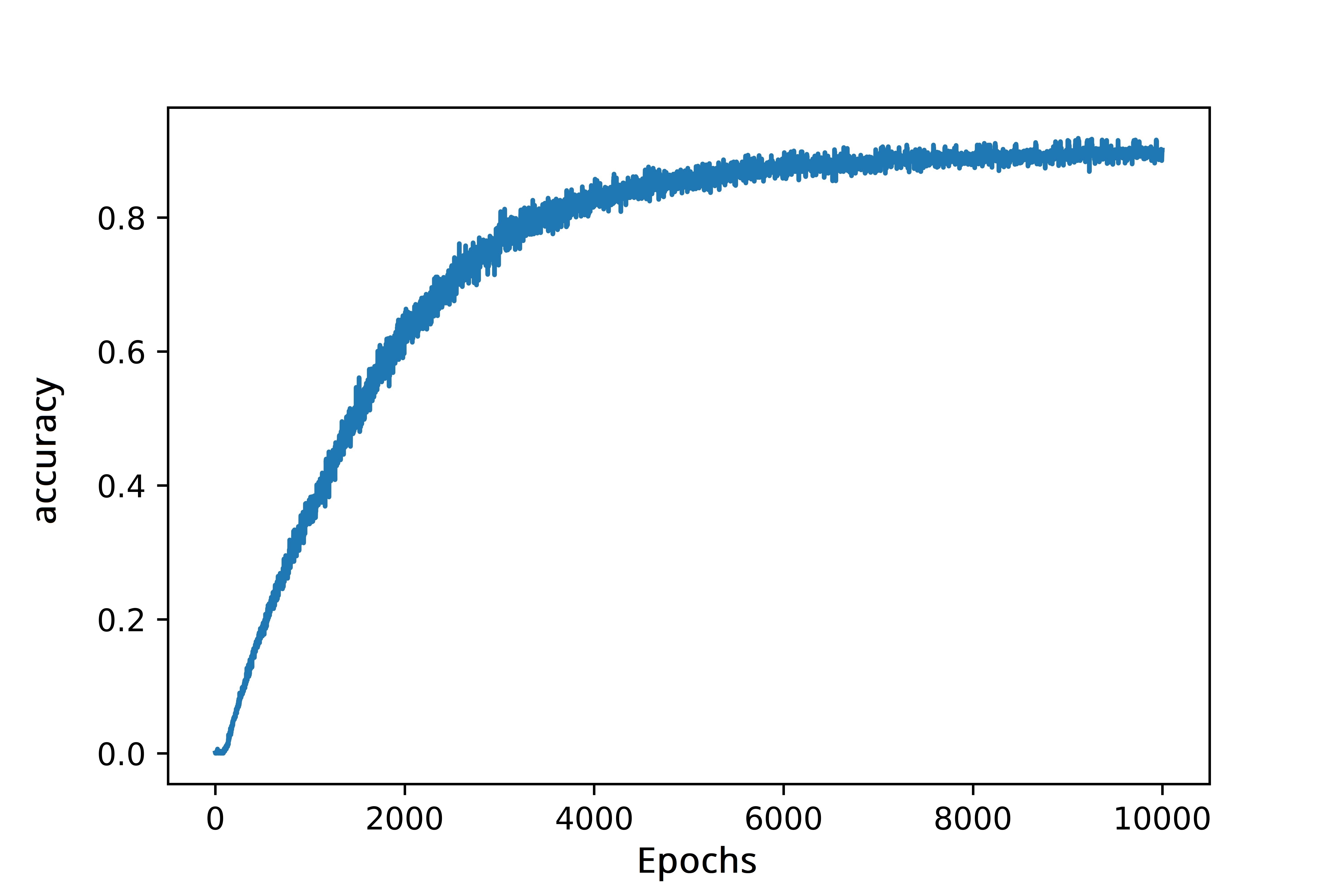}}
	\qquad
	\subfloat[Loss]{\includegraphics[width=7.5cm]{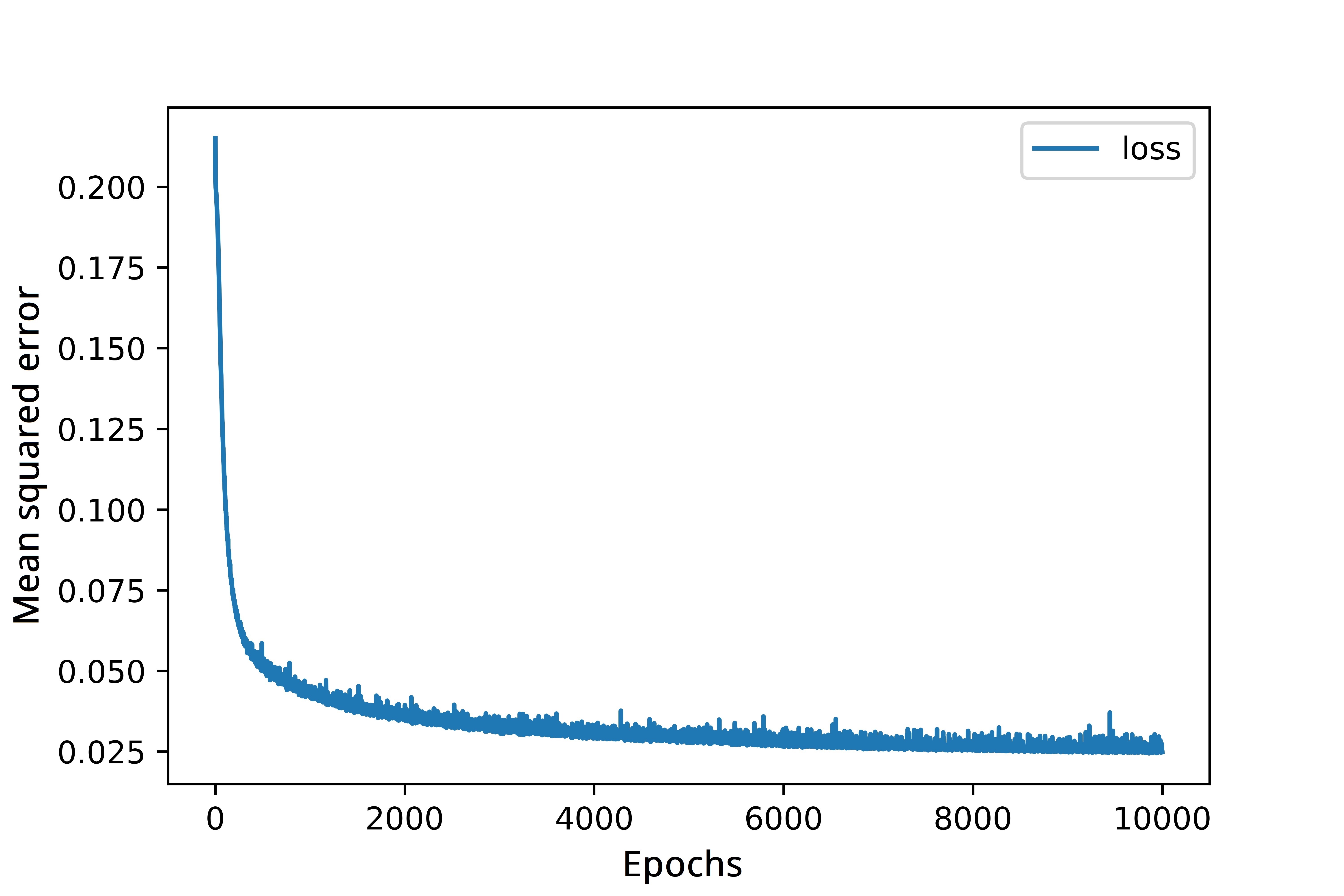}}
	\qquad
	\caption{Curves of a) accuracy and, b) loss function relative to 10000 Epochs for non-restricted network architecture.}
	\label{2fig}
\end{figure*}

\subsubsection{\textit{B. Restricted output}}

In order to increase the learning speed, reduce the number of calculations, and improving the efficiency of a design process, the output of network architecture is restricted in such a way that the DNN should generate the metasurface structure by using the proposed 8 ring-shaped patterns. Unlike the previous approach in which the output generates a 1024 size vector to form the 32×32 metasurface pixels, in this case, the output will generate a 48 size vector. More specifically, each unit-cell consists of 4×4 matrices of these 8 ring-shaped patterns, where each ring-shaped pattern consists of 8×8 pixels. To form the output vector, ring-shaped patterns are denoted by eight digital codes (3-bit) of "000" to "111". Therefore, the output of the DNN generates a 16×3 = 48 size vector.
By restricting the output to produce a 48 size vector, the amount of calculations will be reduced. It will be shown that the accuracy of the network reaches up to 91\%. The details of the designed DNN are summarized in \textbf{Table 2}. The other parameters are similar to the non-restricted output network. \textbf{Fig. 5} shows the curves of the loss function and accuracy. 
\begin{figure*}[h]
	\centering
	\subfloat[Accuracy]{\includegraphics[width=7.5cm]{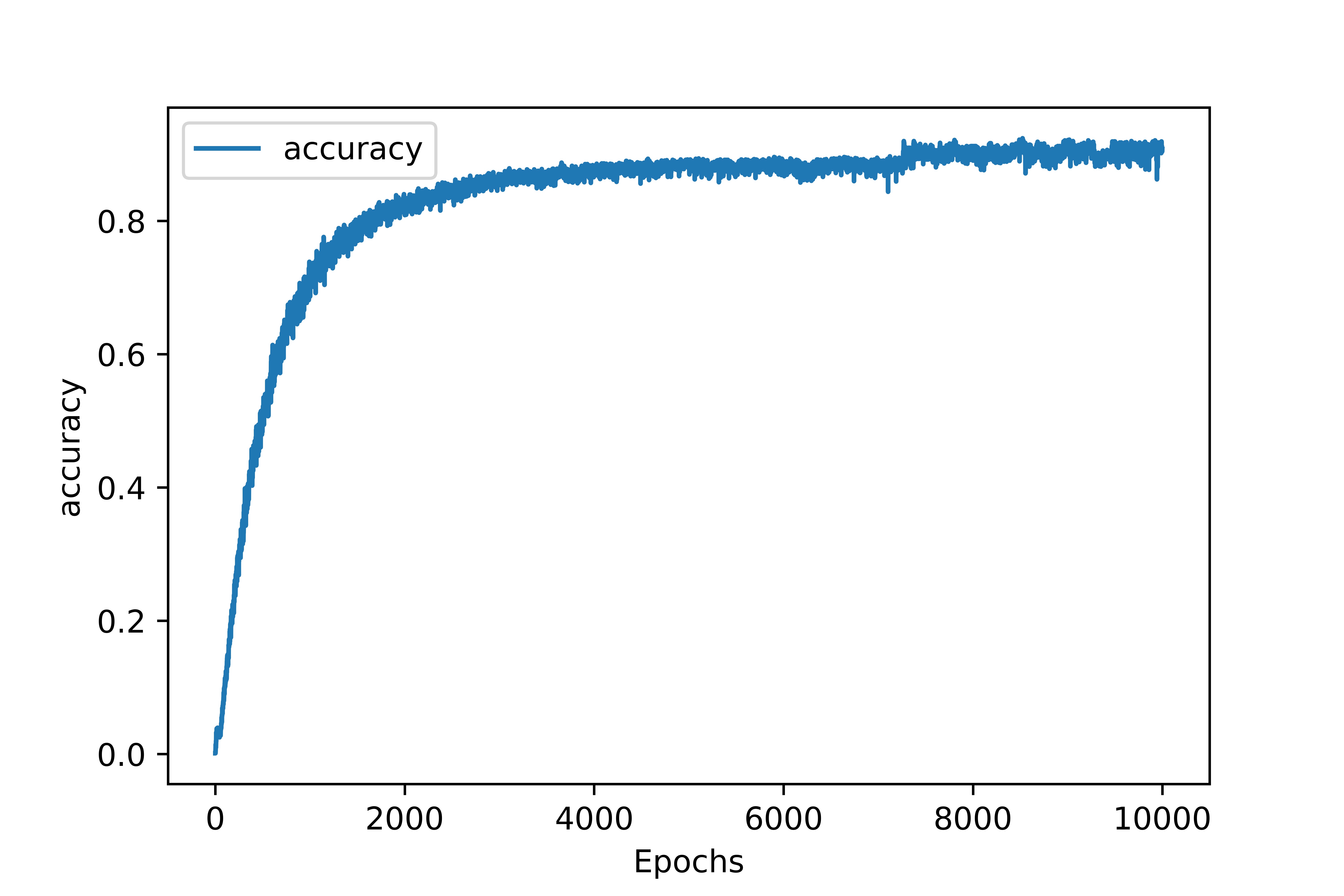}}
	\qquad
	\subfloat[Loss]{\includegraphics[width=7.5cm]{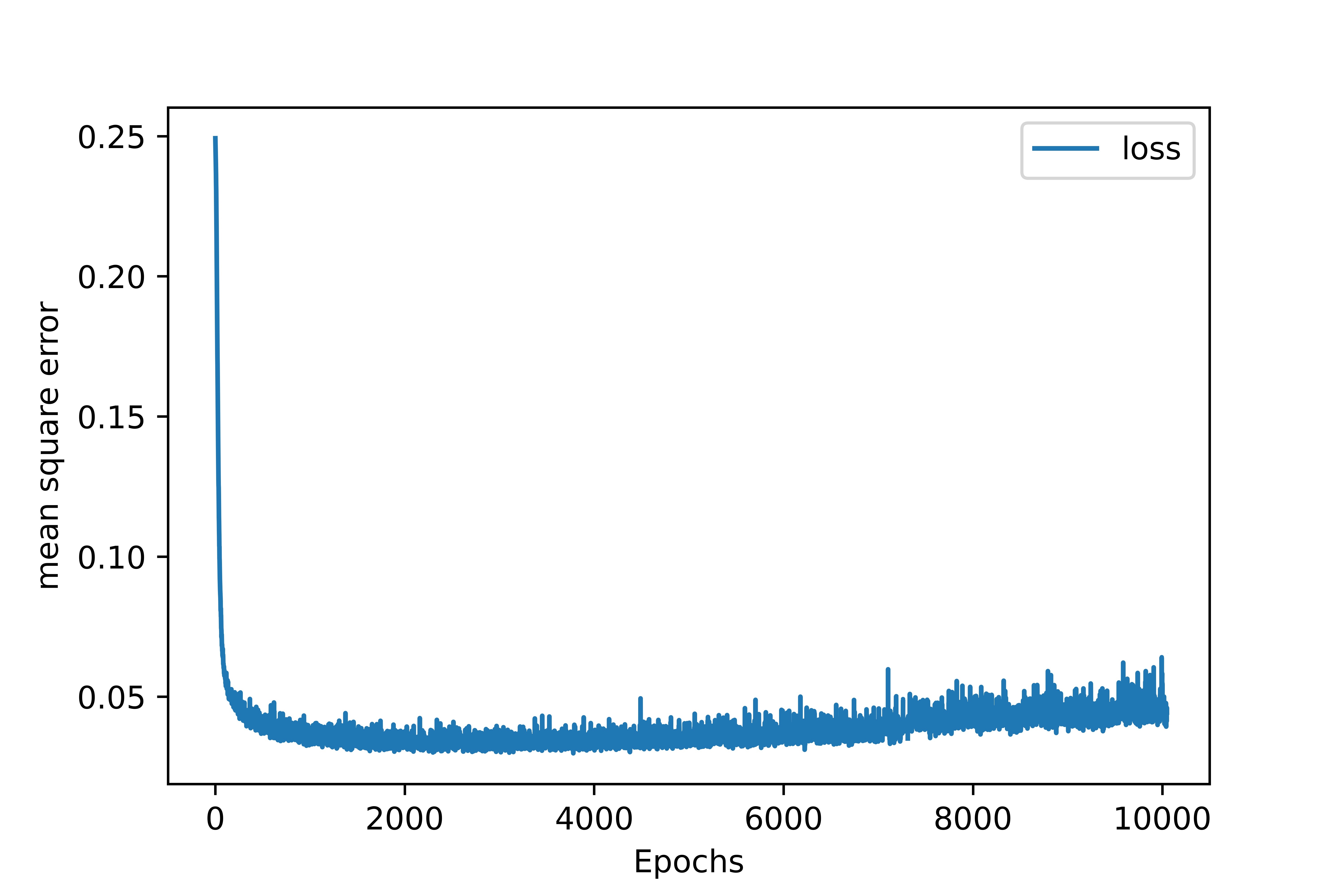}}
	\qquad
	\caption{Curves of a) accuracy and, b) loss function relative to 10000 Epochs for restricted network architecture.}
	\label{2fig}
\end{figure*}

\begin{table}[h]
	\centering
	\caption{Detailed information of the restricted output network architecture}
	\scalebox{0.88}{
		\begin{tabular}{c|c|c|c|c}
			\hline
			Layer number & Layer                & output shape & number of parameter & activation function \\ \hline
			1            & dense\_1 (Dense)     & (None, 24)   & 600                 & relu                \\ \hline
			2            & dropout\_1 (Dropout) & (None, 24)   & 0                   & -                   \\ \hline
			3            & dense\_2 (Dense)     & (None, 500)  & 12500               & relu                \\ \hline
			4            & dropout\_2 (Dropout) & (None, 500)  & 0                   & -                   \\ \hline
			5            & dense\_3 (Dense)     & (None, 500)  & 250500               & relu                \\ \hline
			6            & dropout\_3 (Dropout) & (None, 500)  & 0                   & -                   \\ \hline
			7            & dense\_4 (Dense)     & (None, 500)  & 250500               & relu                \\ \hline
			8            & dropout\_4 (Dropout) & (None, 500)  & 0                   & -                   \\ \hline
			9            & dense\_5 (Dense)     & (None, 500)  & 250500               & relu                \\ \hline
			10           & dense\_6 (Dense)     & (None, 48) & 24048              & sigmoid             \\ \hline
	\end{tabular}}
	
\end{table}

To further validate the effectiveness of the proposed DNN method for restricted output, four different examples are presented. The specified S-parameters are provided into our network and the matrix of unit cells are generated through the input S-parameters. We re-enter these matrices into CST MWS to simulate the reflection coefficient of the metasurface. The simulated results are in good accordance with our desired design target (See \textbf{Table. 3} and \textbf{Fig. 6}).

Consequently, it has been amply demonstrated that the proposed DNN method is superior to other inverse design algorithms of metasurface structure, either from the perspective of computational repetitions, teaching time consumption, and network accuracy. The conformity between the simulated results and design targets promises that the proposed DNN approach is an effective method of metasurfaces design for a variety of practical applications.
	\begin{figure*}[h]
	\centering
	\includegraphics[height=8cm]{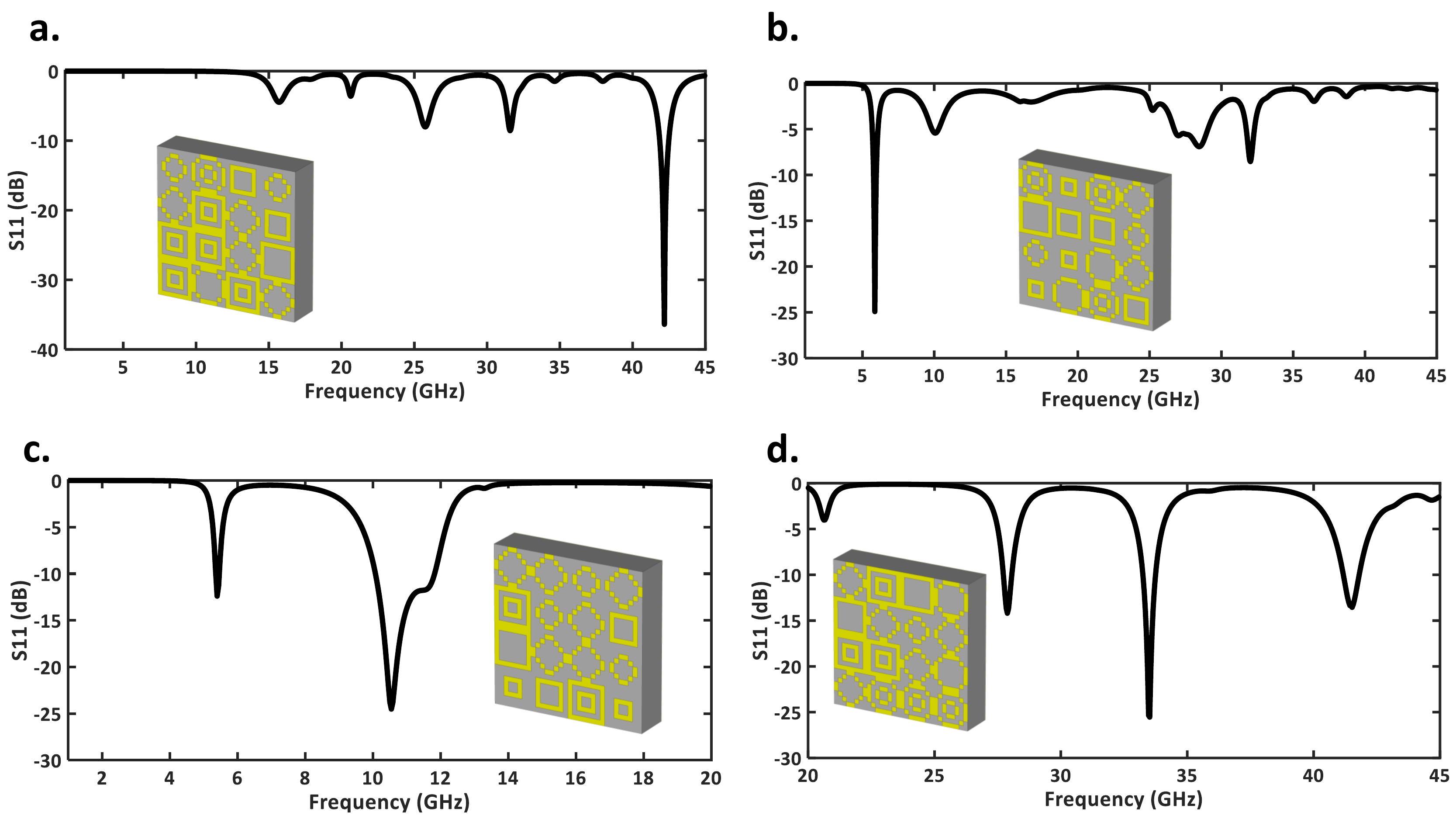}
	\caption{Metasurface design examples through restricted output network architecture.}
	\label{fgr:example2col}
\end{figure*}

\begin{table}[h]
	\centering
	\caption{Desired input targets for four S-parameters which are presented in Fig. 6. }
	\scalebox{0.78}{
		\begin{tabular}{c|c|c|c|c}
			\hline
			Examples & Number of notches                & notches frequency (GHz) & notches depth (dB) & notches bandwidth (GHz) \\ \hline
			Fig. 5a            & 1     & 42    &-35  & 0.7                \\ \hline
			Fig. 5b            & 1 & 5.8  & -25                    & 0.2                   \\ \hline
			Fig. 5c            & 2 & 5.5, 10.5   & -12.5, -24.5              & 0.1, 1.8                \\ \hline
		Fig. 5d            & 3 & 28, 33.5, 41.5  & -14, -25, -13.5 & 0.3, 0.5, 0.7                  \\ \hline
	
	\end{tabular}}
	
\end{table}

\section{Conclusion}
Herein, we have proposed an inverse metasurface design method based on a deep neural network, whereby metasurface structures may be computed directly by merely specifying the design targets. After training the deep learning model by a set of samples, our proposed model can automatically generate the metasurface pattern as the output by four specified reflection criteria (namely, number of resonances, resonance frequencies, resonance depths, and resonance bandwidths) as the input in an ultra-wide operating frequency. Comparing the numerical simulations with the desired design target illustrates that our proposed approach successfully generates the required metasurface structures with an accuracy of more than 90\%. By using 8 ring-shaped patterns during the training process, restricting the output of the network to generate a 48 size vector, our presented method serves as a fast and effective approach in terms of computational iterations, design time consumption, and network accuracy. The presented DNN-based method can pave the way for new research avenues in automatic metasurface realization and highly complicated wave manipulations.


\section{Conflict of Interest}
The author declare no conflict of interest.
\bibliography{REF}


\end{document}